\documentclass{sigchi}

\toappear{Permission to make digital or hard copies of part or all of this work for personal or classroom use is granted without fee provided that copies are not made or distributed for profit or commercial advantage and that copies bear this notice and the full citation on the first page. Copyrights for third-party components of this work must be honored. For all other uses, contact the Owner/Author. \\ 
Copyright is held by the owner/author(s). \\
{\emph{CSCW '17}}, February 25 - March 01, 2017, Portland, OR, USA \\
ACM 978-1-4503-4335-0/17/03. \\
http://dx.doi.org/10.1145/2998181.2998232}

\usepackage[utf8]{inputenc}
\usepackage{multirow}
\usepackage{balance}  
\usepackage{graphics} 
 \usepackage[T1]{fontenc}
\usepackage{txfonts}
\usepackage{times}    
\usepackage{color}
\usepackage{textcomp}
\usepackage{booktabs}
\usepackage{threeparttable}
\usepackage{csquotes}
\usepackage{enumitem}
\usepackage{framed}
\usepackage[dvipsnames]{xcolor}

\def\pprw{8.5in}
\def\pprh{11in}

\definecolor{linkColor}{RGB}{6,125,233}
\usepackage[breaklinks]{hyperref}
\hypersetup{%
  pdftitle={SIGCHI Conference Proceedings Format},
  pdfauthor={LaTeX},
  pdfkeywords={SIGCHI, proceedings, archival format},
  bookmarksnumbered,
  pdfstartview={FitH},
  colorlinks,
  citecolor=black,
  filecolor=black,
  linkcolor=black,
  urlcolor=linkColor,
  breaklinks=true,
}

\makeatletter
\def\url@leostyle{%
  \@ifundefined{selectfont}{\def\UrlFont{\sf}}{\def\UrlFont{\small\bf\ttfamily}}}
\makeatother
\begin{document}

\title{Black Lives Matter in Wikipedia: Collaboration and Collective Memory around Online Social Movements}

\numberofauthors{3}

\author{%
 \alignauthor{Marlon Twyman\\
       \affaddr{Technology \& Social Behavior}\\
       \affaddr{Northwestern University}\\       
       \affaddr{Evanston, IL, USA}\\
       \email{marlont2@u.northwestern.edu}}
 \alignauthor{Brian C. Keegan\\
       \affaddr{Information Science}\\
       \affaddr{Univ. of Colorado, Boulder}\\
       \affaddr{Boulder, CO, USA}\\
       \email{brian.keegan@colorado.edu}}
 \alignauthor{Aaron Shaw\\
       \affaddr{Communication Studies}\\
       \affaddr{Northwestern University}\\
       \affaddr{Evanston, IL, USA}\\
       \email{aaronshaw@northwestern.edu}}
}

\maketitle
\begin{abstract}
Social movements use social computing systems to complement offline mobilizations, but prior literature has focused almost exclusively on movement actors' use of social media. In this paper, we analyze participation and attention to topics connected with the \textit{Black Lives Matter} movement in the English language version of Wikipedia between 2014 and 2016. Our results point to the use of Wikipedia to (1) intensively document and connect historical and contemporary events, (2) collaboratively migrate activity to support coverage of new events, and (3) dynamically re-appraise pre-existing knowledge in the aftermath of new events. These findings reveal patterns of behavior that complement theories of collective memory and collective action and help explain how social computing systems can encode and retrieve knowledge about social movements as they unfold.
\end{abstract}

\keywords{Social movements; civil rights; computer-supported collective action; collaboration network; social computing; user behavior modeling}

\category{H.5.3}{Information Interfaces and Presentation}{Group and
Organization Interfaces (collaborative computing, computer
supported cooperative work)} 
\category{K.4.2}{Computers and Society}{Social Issues} 

\section{Introduction}
Contemporary social movements use social computing and social media platforms to mobilize supporters, negotiate meaning, alter their relationship to media gatekeepers, and reframe issues~\cite{bennett_logic_2013, meraz_networked_2013}. How does knowledge about a social movement come into being while the movement unfolds? What dynamics characterize participation in social computing systems around ongoing and contentious political events as a movement coalesces? 

This paper analyzes how the \textit{Black Lives Matter} movement ($BLM$) has manifested itself within the English language Wikipedia. The English Wikipedia is the largest language edition and includes articles on topics relevant to the current study: breaking news, social movements, protests, and social justice issues. Wikipedia is also a place for collective memory building around historical events, such as the Arab Spring and the Vietnam War~\cite{ferron_collective_2011, luyt_wikipedia_2015, pentzold_fixing_2009}, as well as unfolding coverage of breaking news~\cite{keegan_hot_2011, keegan_hot_2013, keegan_history_2013}. 
$BLM$ provides a case study of a social movement generating breaking news around social justice issues across the United States.

We use trace data from $BLM$ relevant articles' revision histories and pageviews on Wikipedia to understand dynamics of attention, collective memory, and knowledge production in response to a movement. Prior work on social movements and social computing has focused disproportionately on Twitter and Facebook use, emphasizing interpersonal interactions, event coordination, and information diffusion dynamics among movement participants. In contrast, Wikipedia exemplifies a collaborative knowledge and memory production system that aims to remain independent of movement influence. Analyzing the temporal changes around movement-related articles offers an ideal perspective on how knowledge about a social movement shifts while it is still in action. 

Our results complement and extend existing analyses of social computing activity around social movements as well as prior work on Wikipedia participation dynamics. 
We find that movement events and external media coverage appear to drive attention and editing activity to $BLM$-related topics on Wikipedia. This activity and attention intensified as the movement gained prominence and momentum, leading to bigger spikes of attention to movement events, faster coverage of new movement topics, expanded discussion, and the emergence of networks of attention and collaboration around $BLM$-related articles. Specifically, we describe three key processes that we observe in Wikipedia around topics related to $BLM$: intensified documentation, collaborative migration, and dynamic re-appraisal.

Our findings show novel patterns of collective memory that occur in social computing in conjunction with social movement events. In addition, we describe novel dynamics of attention and collaboration within Wikipedia while also replicating patterns from prior Wikipedia research in other topic domains. We look only at events connected to a single movement, but the processes we observe may help explain general patterns of collective memory making within social computing systems.

\section{Background and Prior Work}
\subsection{The ``Black Lives Matter'' movement}
$BLM$ emerged in the United States during the summer of 2014 in response to instances of police officers shooting or killing unarmed African Americans. The roots of the movement stretch much further back in time and, for some supporters, encompass the history of racial violence, exclusion, inequality, mass incarceration, and slavery in the United States. However, most accounts of the movement place its immediate origins in Ferguson, Missouri with the death of Michael Brown on August 9, 2014.

Michael Brown, an 18 year-old African American man, was killed in Ferguson, Missouri by Darren Wilson, a 28 year-old white Ferguson police officer. In response to Brown's death, protesters and activist groups adopted the phrase ``Black Lives Matter'' to metonymize their grievances. This phrase originated in response to the shooting death of another unarmed African American teenager, Trayvon Martin, in Sanford, Florida early in 2012~\cite{graeff_battle_2014}. As it became clear that the events in Ferguson had provoked nationwide concern, the phrase spread on social media and in the press~\cite{freelon_beyond_2016}. Subsequent events from 2014 through 2016 involving the deaths of African Americans associated with police broadened the attention around ``Black Lives Matter'' as a movement focused on racial injustice and police violence. The political climate and ambiguity surrounding $BLM$'s goals has since generated numerous supporters and critics of the movement. 

From the beginning, social media systems played a critical role in the rise and popularization of $BLM$~\cite{graeff_battle_2014}. Twitter has generated a great deal of activity as a place where content and issues specific to the African American community have gained attention~\cite{sharma_black_2013}. Following the deaths of Martin, Brown, and others, the dissemination of images, videos, and hashtags on Twitter and other social media served a pivotal role leading to the emergence of $BLM$~\cite{freelon_beyond_2016, jackson__2015}. However, the impact of the $BLM$ movement in social computing extends far beyond social media. We seek to understand this broader impact by studying activity in Wikipedia around topics related to $BLM$.

\subsection{Social movements \& social computing systems}
Social media platforms have played influential roles in mobilizing participation in social movements. These platforms are designed to reduce coordination costs as well as enhance information sharing, potentially encouraging more people to engage in protest politics than would have otherwise~\cite{tufekci_social_2012}. ``Networked counterpublics'' deploy novel tactics such as ``threadjacking'' to garner attention, recruit newcomers, and mobilize supporters~\cite{jackson_hijacking_2015}. Broadly speaking, these are all instances of \emph{computer-supported collective action}~\cite{zhang_wedo:_2014}. 
We investigate a set of related concerns about how computer-supported collective action unfolds in the context of commons-based peer production system:
what patterns of attention, engagement, and collaboration emerge in social computing systems engaged in knowledge collaboration around contentious movements? 

We extend prior work by investigating participation and attention in a peer production system that was not a primary site of mobilization for the movement under study. Recent studies of $BLM$ have relied heavily on data from social media to examine how movement participants curate information and mobilize for action~\cite{de_choudhury_social_2016, freelon_beyond_2016, jackson__2015, jackson_hijacking_2015}.
In contrast, the contributors and readers of articles related to $BLM$ on Wikipedia do not necessarily participate in the movement or share the perspectives of movement actors. Thus, rather than focus on social media as a mode of movement participation, we investigate whether and how the patterns of knowledge collaboration and information consumption in an online peer production community shape the representation of a social movement.

\subsection{Breaking news \& current events in Wikipedia}
The Wikipedia collaborations around breaking news display unique patterns of activity and engagement within emergent topic spaces. Editors demonstrate remarkable ability to respond to current events: they create articles within minutes of major disasters and catastrophes~\cite{keegan_hot_2011}, alter collaboration practices to support high-tempo interactions~\cite{keegan_staying_2012}, rapidly regenerate prior organizational structures~\cite{keegan_hot_2013}, and adopt differentiated social roles~\cite{keegan_emergent_2015}. A broader literature has shown that editors act interdependently assuming nuanced social roles~\cite{arazy_functional_2015,welser_finding_2011}, managing conflict dynamics~\cite{kittur_he_2007,yasseri_dynamics_2012}, and coordinating their actions to produce high quality content~\cite{kittur_harnessing_2008}. Other research has approached the study of the Wikipedia community itself through the focus of social movement theory (e.g.,~\cite{jemielniak_common_2014,konieczny_governance_2009}). This prior work has considered the dynamics of breaking news coverage on Wikipedia and has described Wikipedia editing as a form of social movement activity, but has not focused on editing activity around social movement events as they are occurring.

Articles related to $BLM$ in the English language Wikipedia provide an ideal case to understand more about current social movements through the lens of social computing systems. The movement is a current event, the associated deaths and protests made breaking news, and the topic occupies a domain that bridges different content areas in the platform. Collaboratively producing knowledge about social movements could be dynamic and contentious in ways that other breaking news events are not. Prior research on breaking news in Wikipedia does not consider the need to situate new events with the context of older events. $BLM$, in particular, focuses on the deaths of a minority group in the United States, is embedded within historically contentious issues, and continues to generate new information. This case can extend our understanding of breaking news practices in Wikipedia to include a social movement and related events. \\  

\vspace{1.2em} 
\begin{framed}
\textbf{RQ1}: How has Wikipedia editing activity and the coverage of $BLM$ movement events changed over time?
\end{framed}

\vspace{1.2em} 
\begin{framed}
\textbf{RQ2}: How have Wikipedians collaborated across articles about events and the $BLM$ movement?
\end{framed}

\begin{figure*}[tb!] 
    \centering
    \includegraphics[width=1.8\columnwidth]{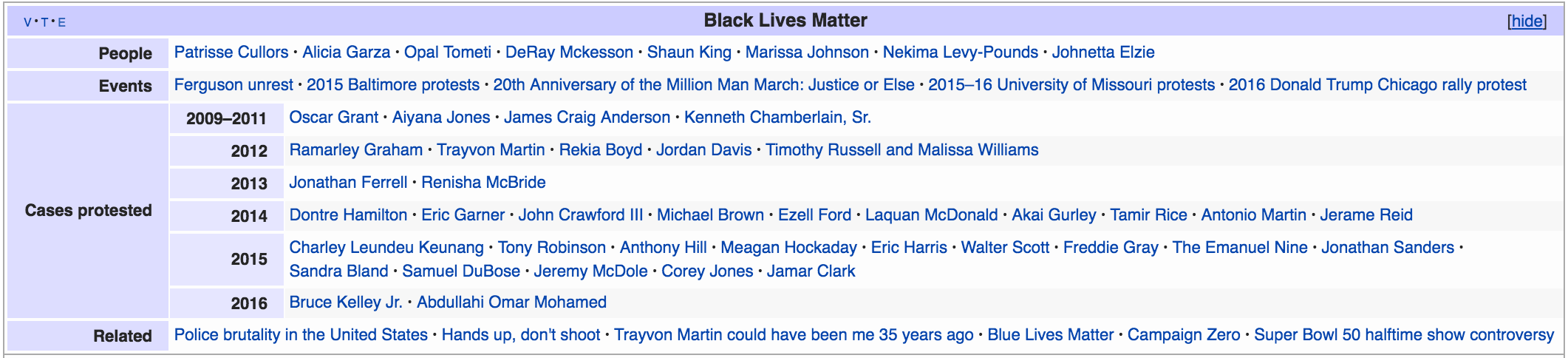}
    \caption{The Wikipedia Template:Black Lives Matter containing relevant articles for analysis.}
    \label{template_fig} 
\end{figure*}

\subsection{Collective memory in social computing}
$BLM$ also lets us explore how social computing systems serve as sites of collective memory. The movement centers around a series of recent deaths. Social movements often focus on transforming individual events such as these into collective memories~\cite{harris_it_2006,schuman_generations_1989}. A groups' shared sense of the past contributes to collective action by providing common frames to articulate grievances, develop shared identity, regenerate networks and resources, and increase incentives for cooperation~\cite{gongaware_collective_2010,mcadam_recruitment_1986}. Halbwachs's theory of collective memory argues that memories shape and are shaped by present concerns and developed through social interactions~\cite{halbwachs_collective_1992}. Observing memory building around a social movement with respect to related events provides new insights into collective memory in Wikipedia.

Prior work by Pentzold~\cite{pentzold_fixing_2009} describes the role of Wikipedia in constructing collective memories. Wikipedia's openness, popularity, and detailed archival records enables investigations into the processes of translating a topic's archival materials, historical scholarship, and other memories into an article. Subsequent research followed this perspective with a number of case studies. Memory-building processes for the Egyptian revolution across languages~\cite{ferron_collective_2011}, the deaths of notable people~\cite{keegan_is_2015}, and the Vietnam War~\cite{luyt_wikipedia_2015} have all extended the understanding of how editors collaborate, manage conflicts, and produce knowledge in a commemorative mode. The current study extends the application of collective memory to events and memories encoded into individual articles that in turn compose a larger network documenting evidence relevant to a social movement. 

However, there are opposing perspectives regarding whether collective memory threatens or supports collective action. Each perspective implies different dynamics between activity and attention across current and historical events. On one hand, tragic events in the past act as proof that transgressing the status quo is dangerous, which reinforces shame, fear, and apathy~\cite{salamon_fear_1973}. Such events can be collectively self-censored and more favorable events privileged in order to preserve a positive self-image or dominant social identity. On the other hand, tragic events can be translated and sustained in oral, written, or visual histories as well as commemorated through rituals (\textit{e.g.}, anniversaries) and artifacts (\textit{e.g.}, literature). These collective memory repertoires provide interpretive material for people to make sense of current events, which in turn support solidarity and collective action~\cite{harris_it_2006}. In this view, activity around current events should be positively correlated with historical events as they are re-surfaced and sustained to support information seeking and mobilization.

\vspace{.4em}
\begin{framed}
\textbf{RQ3}: How are events on Wikipedia re-appraised following new events?
\end{framed}

\section{Data and Methods}
\begin{table*}[!htbp] 
\begin{center}
\begin{threeparttable}
    \begin{tabular}{@{\extracolsep{5pt}} @{}lrrrrr@{}} 
    \toprule
    Article & Revisions & Editors & Talk Revisions & Talk Editors & Pageviews \\ 
    \hline \\[-1.8ex] 
    Shooting of Michael Brown & $7,141$ & $904$ & $19,454$ & $391$ & 4,254,371 \\ 
    Shooting of Trayvon Martin & $6,639$ & $805$ & $19,707$ & $782$ & 1,145,483 \\ 
    Shooting of Oscar Grant & $2,728$ & $756$ & $985$ & $113$ & 1,066,511\\ 
    Charleston church shooting & $2,544$ & $575$ & $2,118$ & $162$ & 903,191 \\ 
    Ferguson unrest & $1,847$ & $556$ & $1,640$ & $139$ & 1,315,598\\ 
    Black Lives Matter & $1,626$ & $408$ & $534$ & $74$ & 450,516\\ 
    Death of Eric Garner & $1,526$ & $486$ & $1,178$ & $122$ & 1,873,056\\ 
    Death of Freddie Gray & $1,513$ & $354$ & $1,358$ & $115$ & 1,375,498 \\ 
    2015 Baltimore protests & $1,034$ & $298$ & $634$ & $248$ & 611,371\\ 
    Death of Sandra Bland & $950$ & $284$ & $1,012$ & $65$ & 492,589\\ 
    \hline \\[-1.8ex] 
    Total & $27,548$ & $4,372$\tnote{*} & $48,646$ & $1,743$\tnote{*} & 13,488,184 \\
    \bottomrule
        \end{tabular}
    \begin{tablenotes}
    \item[*] \footnotesize{Indicates a unique count for each measure. Overall, $5,449$ unique editors contributed.\\}
    \end{tablenotes}
    \caption{List of articles with the most revisions in the sample related to the movement. The number of editors, talk page activity, and pageviews are also included.} 
  \label{tab:rev_activity} 
\end{threeparttable}
\end{center}
\end{table*} 
This study uses the revision histories of English language Wikipedia pages related to $BLM$ collected on April 21, 2016. A custom Python script built with the Wikitools library\footnote{https://github.com/alexz-enwp/wikitools} retrieved data from the English Wikipedia's MediaWiki API.\footnote{https://en.wikipedia.org/w/api.php} The creation of the analysis page set required multiple steps. First, we built a list of potential pages by identifying the 59 articles belonging to the \textit{Black Lives Matter} Wikipedia template (Figure~\ref{template_fig}) and the 74 articles in the \textit{Black Lives Matter} Wikipedia category. From the union of these unique pages, we excluded pages if they were not focused on the movement, a death, or a protest event. These pages were excluded because of the current study's focus on coverage and attention related to the movement, motivating deaths, and the protests in response. The resulting articles covered the social movement itself (1 page), deaths of African Americans (37 pages), and protest events (4 pages) totaling 42 pages. 

We included the redirects and talk pages as well. Redirects give insights into editors who were interested in contributing, but may have edited a similar page before the current version became the main article for a topic~\cite{hill_consider_2014}. We map activity on the redirected page to the target article of the redirect. The overall editing on redirect pages represents a small amount of activity to the sample (1.7\% of total), but is included to provide article creation data. Talk pages are a history of discussions that occur throughout the lifespan of articles. In the complete set, there were 86,940 revisions made by 6,795 editors to 141 articles and talk pages from January 6, 2009 to April 21, 2016. 

To supplement the revision data, we also analyze the pageviews for the sample from January 1, 2014 to January 1, 2016. Pageviews are a count of the number of visitors to each Wikipedia article. The data was extracted using Python from a third-party database of daily pageviews data.\footnote{http://stats.grok.se/} While this time range truncates some of the information that can be gained for some of the article sample, the two years of pageviews shows information consumption patterns about articles in the sample and daily changes in attention, especially after the movement began.  

We focus on the three aforementioned research questions in order to extend prior research on social computing, social movements, and collective memory as well as studies of participation around breaking news and current events in Wikipedia. $BLM$ provides a unique and exceptional opportunity to advance research at the intersection of these topics. Because prior work addressing these concerns and related phenomena does not support clear predictions or hypotheses to test, we adopt an exploratory and descriptive approach. We gain insights into the editing activity, attention, and collaboration surrounding $BLM$ on Wikipedia. Temporal dynamics, correlations, and networks uncover different behavioral patterns showing the movement's impact on Wikipedia. Additionally, we qualitatively review discussions from the movement's talk page to further investigate collaborations surrounding $BLM$.

\section{Results}
\subsection{RQ1: Intensified Documentation}
\begin{figure}[tb]
    \centering
    \includegraphics[width=\columnwidth]{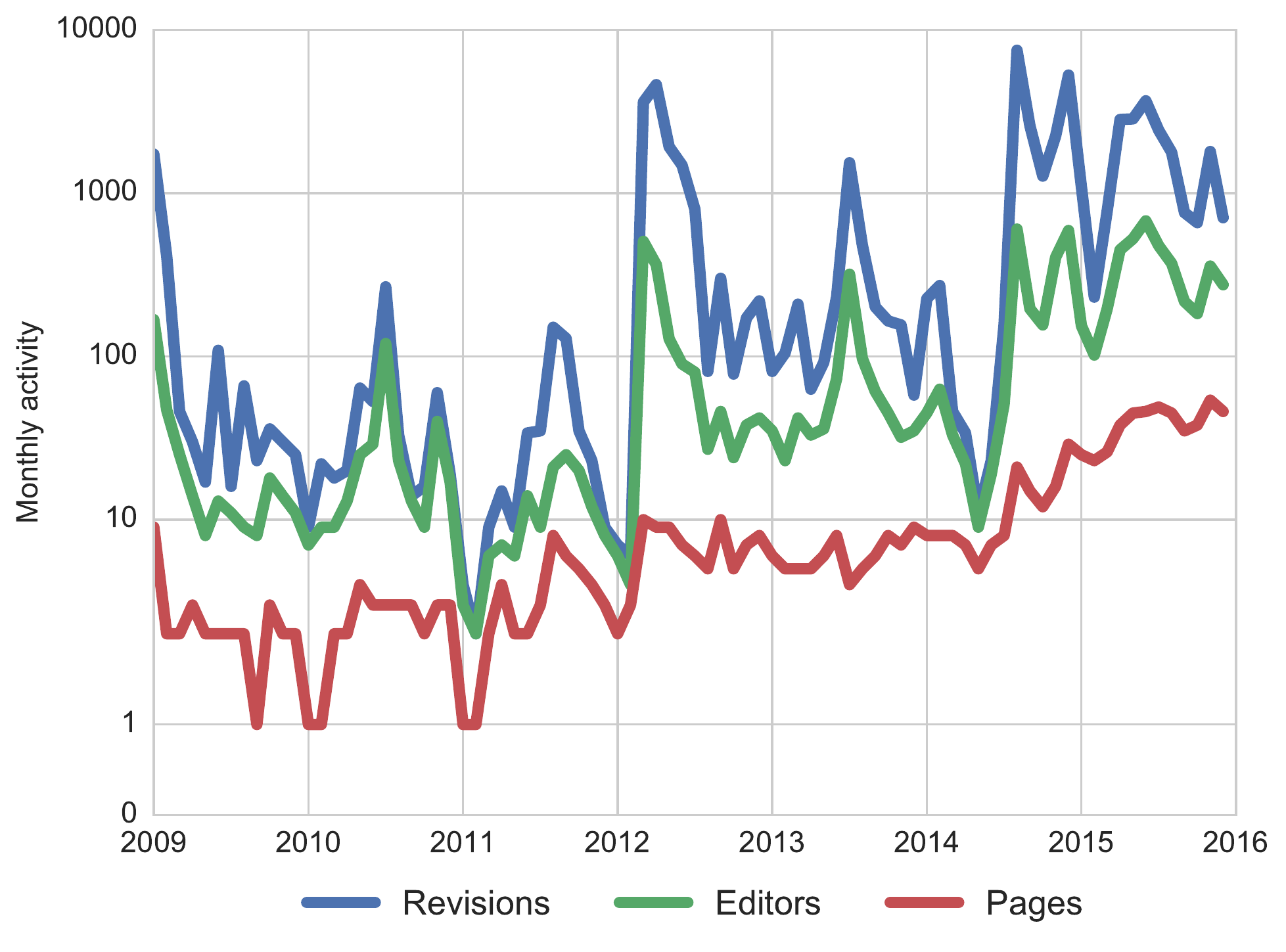}
    \caption{Aggregated monthly activity over time for all the pages included in analysis. The number of revisions (blue line), editors (green), and pages (red) are plotted.}
    \label{fig:freq_plot}
\end{figure}
\begin{figure*}[tb] 
    \centering
    \includegraphics[width=1.6\columnwidth]{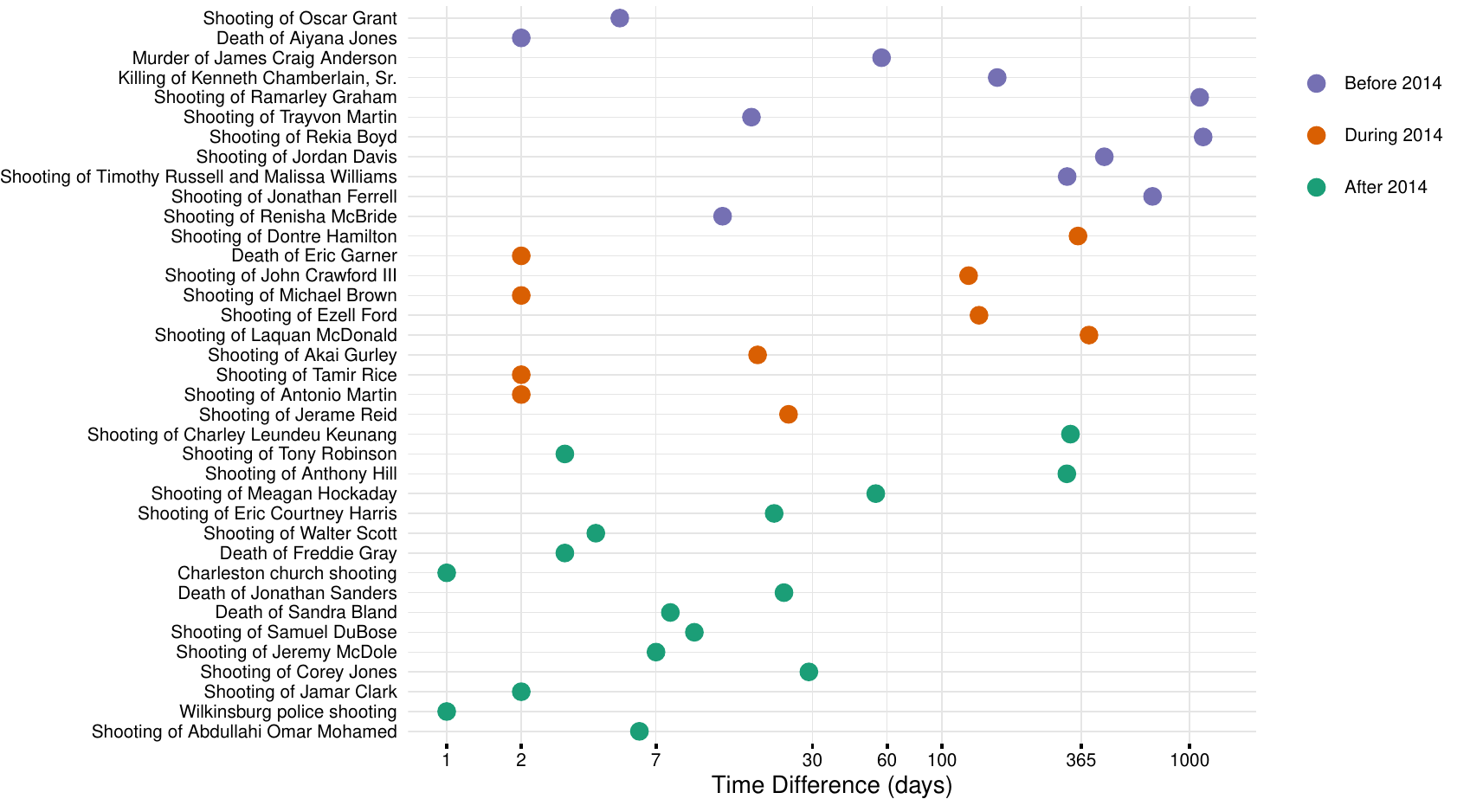}
    \caption{The time difference (lag) between $BLM$-related deaths and the date that the corresponding Wikipedia page is created. The $y$-axis is ordered chronologically from oldest death (top) to most recent. Points for each death are grouped by color corresponding to when they occurred: ``Before 2014'', ``During 2014'', or ``After 2014''.}
    \label{fig:timedelay} 
\end{figure*}
Research Question 1 asked, ``How has Wikipedia editing activity and coverage of the $BLM$ movement events changed over time?'' We examine this question in two ways. First, we describe the revision and pageview activity to the top articles and identify variation in monthly activity due to current events. Second, we observe changes in the latency between the occurrence of an event and the creation of its article. There is wide variance, but latencies decline for events since the beginning of $BLM$. These results suggest Wikipedia editors engage in an emergent practice we call \textit{intensified documentation} as they create more content, more quickly, within the normative constraints of Wikipedia policies like neutral point of view or notability. As $BLM$ gained notoriety, Wikipedians covered new events more rapidly and also created articles for older events.

Revision histories document several dimensions of knowledge creation in Wikipedia: the editor, time stamp, and content changed in each revision. Table~\ref{tab:rev_activity} shows the number of revisions, editors, talk page revisions, talk page editors, and pageviews to the ten articles with the most revisions. The amount of activity on these ten articles and talk pages dominates all of the activity on the other 121 pages in the sample combined. The Top 10 pages and their corresponding talk pages account for 76,194 (87.6\%) of all revisions, and 5,449 (80.2\%) editors made these revisions. Excluding $BLM$, these top articles are about events that generated a large amount of media attention outside of Wikipedia~\cite{freelon_beyond_2016,graeff_battle_2014,jackson__2015}. The activity for the ``Shooting of Michael Brown'' and the ``Shooting of Trayvon Martin'' contribute 52,941 (60.9\%) of all revisions by themselves. The ``Shooting of Michael Brown'' article also had the most pageviews, accounting for 31.5\% of all views in the Top 10, reflecting its influence as a major news event. 

Figure~\ref{fig:freq_plot} visualizes the monthly activity for all articles in our corpus by number of revisions made (blue), unique editors contributing (green), and pages edited (red). The peaks in activity correspond closely to the death of Oscar Grant (January 1, 2009), the death of Trayvon Martin (February 26, 2012), the death of Michael Brown (August 9, 2014), the death of Tamir Rice (November 22, 2014), the protests in Baltimore (April 2015), and the Charleston Church shooting (June 17, 2015). We note one peak in July 2013 corresponds to the acquittal of George Zimmerman, the man who killed Martin, and consists almost entirely (98\%) of edits to pages related to the shooting of Trayvon Martin and a few edits to pages about the shooting of Oscar Grant. The monthly activity plots illustrate that prominent events drive periods of high activity and reshape activity in aggregate. While they do not sustain the levels of peak activity, we observe a general trend towards an increasing level of activity across all three metrics as additional articles are created and added to the sample. However, the large peaks seem to suggest that activity in the $BLM$ topic space is focused on individual events and does not necessarily imply sustained writing about the $BLM$-related topics.  

The activity analysis above does not discriminate between different types of editors who may contribute to articles. The editing environment of Wikipedia is dynamic and will be different for different types of users (\textit{e.g.}, registered, administrators, bots, and unregistered). Page protection statuses affect editing activity from different sets of users, especially unregistered ones~\cite{hill_page_2015}. There are 374 log events collected for the sample, including redirected pages. 21 of the 374 events are autoconfirmed page protections to the 42 main articles, which prevent unregistered users from making revisions. In the entire 141-page sample, there were 3,275 unregistered editors who made 7,703 revisions. These people would not have been able to edit anything when pages were under protection. It is possible that these protection spells suppressed activity from unregistered users who were vandalizing pages, and also users interested in contributing, but unfamiliar with the policies of Wikipedia to navigate the protections.

\subsubsection{Temporal dynamics of article creation}
The template for $BLM$ includes 37 incidents that resulted in the deaths of African Americans (Figure~\ref{template_fig}), but how long after these incidents did it take for Wikipedia articles to be generated? The time between an event and the creation of its Wikipedia article event can signal its significance~\cite{keegan_hot_2013}. Unlike the ``wiki-bituary'' activity following the deaths of people who already had Wikipedia articles~\cite{keegan_is_2015}, the deceased subjects in our sample did not have articles until after their deaths generated media attention. Figure~\ref{fig:timedelay} plots the 37 events and the time elapsed between the event and article creation. 

We note the wide variation in article creation latency. Only two of the deaths have articles created within 24 hours of the event (``Charleston Church Shooting'' and ``Wilkinsburg police shooting''); across the rest of events, it takes at least 48 hours. Even the deaths that were major news stories in the United States, such as ``Shooting of Michael Brown,'' ``Shooting of Trayvon Martin,'' and ``Death of Eric Garner'' took at least two days for their articles to be created. Many of the articles appear months or even years after the deaths occurred. The ``Shooting of Rekia Boyd'' (1,133 days) and the ``Shooting of Ramarley Grant'' (1,097 days) had the longest article creation lag. In total, 12 of the events had article creation lags of at least 100 days. None of these events were among the most active pages in the sample by number of revisions. 

Also in Figure~\ref{fig:timedelay}, we separate the deaths into three categories that correspond to different phases of the $BLM$ movement: ``Before 2014,'' ``During 2014,'' and ``After 2014.'' There are 11 death events that took place before 2014, 10 events during 2014, and 16 events after 2014. Before 2014 and the emergence of $BLM$, the 11 articles about deaths took on average 361 days to be created after each death. During 2014, when the $BLM$ movement gained attention, the 10 pages about deaths took on average 107 days to be created. After 2014, the 16 pages about deaths took on average 51 days to be created after each death. The deaths of most of these African Americans did not become articles until after widespread attention started to be given to $BLM$ and these types of incidents. 

The $BLM$ movement has led to both reduced latency in article creation and increased coverage of relevant deaths, demonstrating an overall pattern of \emph{intensified documentation}. The rise of $BLM$ has coincided with Wikipedia reducing its average response time from 361 to 51 days (86\% reduction) when creating new articles about such related deaths. In addition, there have been more articles created about related deaths in the 16 months after 2014 than in the 60 months before 2014. The $BLM$ movement and the curation of a Wikipedia template and category organized the topic space and appears to have helped editors create knowledge about the events related to the movement. 

\subsection{RQ2: Collaborative migration}
\begin{figure}[tb]
    \centering
        \includegraphics[width=\columnwidth]{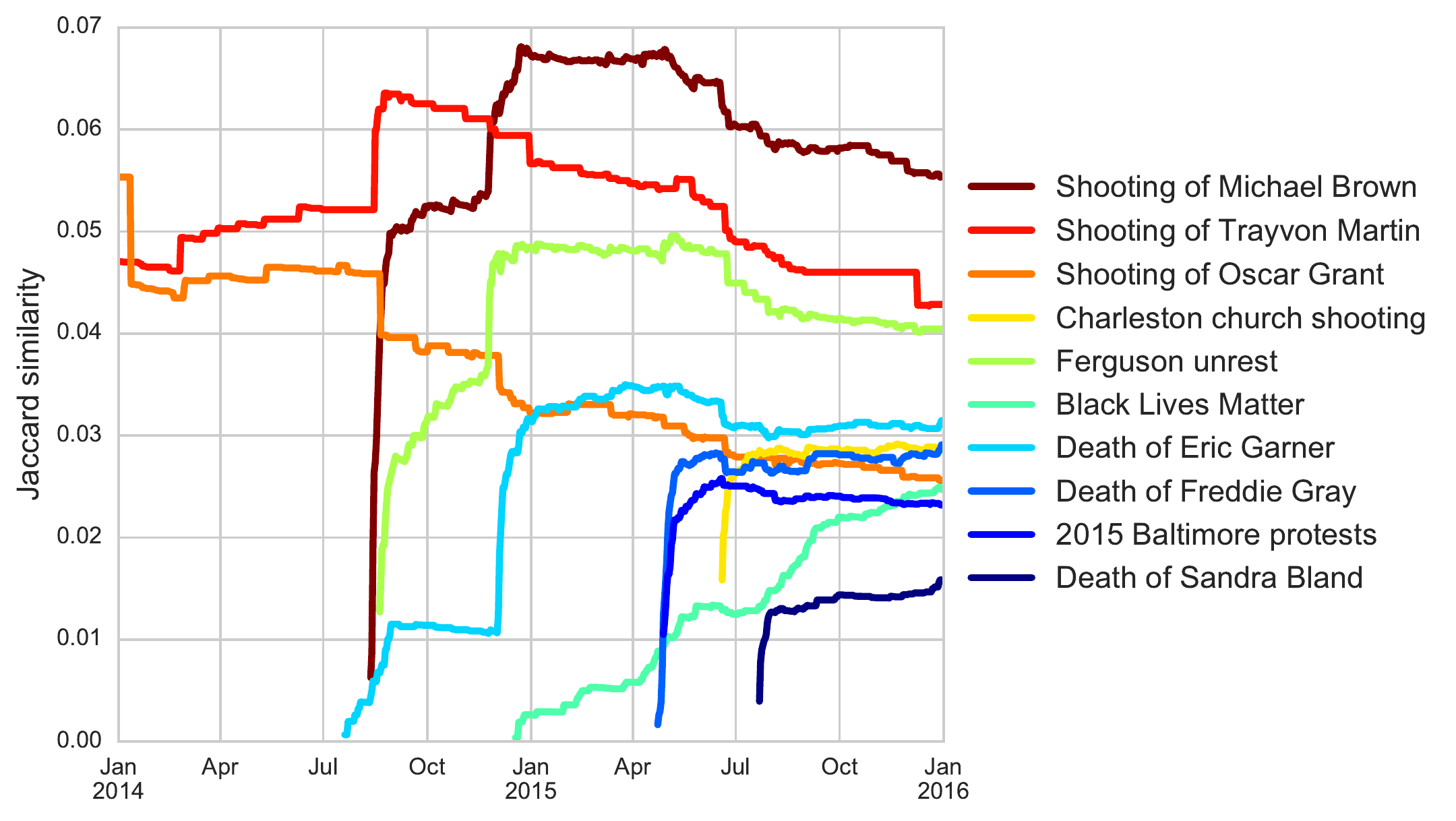}
        \caption{Changes in Jaccard similarity scores for top 10 articles. Similarity is computed daily for the cumulative editor sets between the focal article and editors to all articles in the corpus.}
      \label{fig:jaccard_score}
      \end{figure}
Research Question 2 asked, ``How have Wikipedians collaborated across articles about events and the $BLM$ movement?'' To what extent did Wikipedia editors contribute across related articles? How does this activity change over time and in response to new events? We explore these questions in two ways. First, we examine the similarities between the sets of editors and changes over time across articles. Then we qualitatively investigate discussions on the ``Black Lives Matter'' talk page. Experienced editors were diligent in maintaining a neutral point of view, improving the sourcing, and making decisions about inclusion of related events in the article. These findings illustrate a distinctive mechanism we term \textit{collaborative migration} in which groups of Wikipedia editors move among related event articles in the absence of traditional coordination structures to support co-authorship, such as templates or WikiProjects. In the case of $BLM$, many of these structures were not created until mid-2015.

\subsubsection{Editor similarities over time}
We analyze editor similarity across articles using Jaccard coefficients. The Jaccard coefficient captures the similarity between two sets of entities by computing the size of the intersection between entities in both sets, divided by the union of the sets. We computed daily Jaccard coefficients between the cumulative set of editors to a given article and the cumulative set of editors to all other articles in the corpus between January 2014 and January 2016. Substantively, this metric describes what fraction of users on an article in a given day had edited any other article in the corpus at any time leading up to and including that day. Small values indicate a small percentage of the editors on the article have engaged with other $BLM$-related articles while large values indicate there is substantial overlap between the editors of the article and the editors contributing across all other $BLM$ articles. Assessing the similarity in editors between articles reveals whether the collaborators converge or diverge over time.

Figure~\ref{fig:jaccard_score} plots these daily Jaccard similarity coefficients for the Top 10 articles (Table~\ref{tab:rev_activity}). First, the daily editor similarities between articles are relatively low in absolute terms: the ``Shooting of Michael Brown'' article shows the greatest similarity in editor membership with the rest of the corpus but never rises above 7\% similarity. Given the number of articles and the tendency for most editors to contribute only once, $\sim$7\% similarity over a set of 6,795 unique users is quite high. 

Many newly-created articles like ``Shooting of Michael Brown'' follow a ``rotated L''-shaped pattern of sharply rising similarity immediately following their creation followed by a plateau of relatively constant similarity in the cumulative editor sets. The immediate increase in similarity suggests a set of ``early responder'' editors rapidly come together to contribute to and frame these articles. However, the tendency for editor similarity to decrease over time across these articles suggests that new contributors to these articles in the weeks and months after the events themselves tend to edit only that article rather than working across articles. Figure~\ref{fig:jaccard_score} also shows discontinuities in articles' similarities around the time that \textit{new} articles are created. For some events and some articles, this results in increased similarity scores. For others, the same event appears to drive down similarity, indicating that the article's editors overlap less with the full set.

The creation of the ``Black Lives Matter'' article (Fig.~\ref{fig:jaccard_score}, cyan line) in late 2014 corresponds to noticeable increases in the similarity coefficients for the articles about Michael Brown, Ferguson, and Eric Garner, but minor decreases in the similarity coefficients for Trayvon Martin and Oscar Grant. Editors on the ``Black Lives Matter'' article also started editing the former articles, but did not edit the latter articles. All of the related event articles show a general tendency towards slowly-declining similarity over time. The $BLM$ article's similarity coefficient, while relatively low compared to major event articles in absolute terms, has followed a different and constantly increasing trajectory. The trend suggests editors of event articles increasingly make contributions to the movement article as well.

\begin{figure}[tb]
    \centering
        \includegraphics[width=\columnwidth]{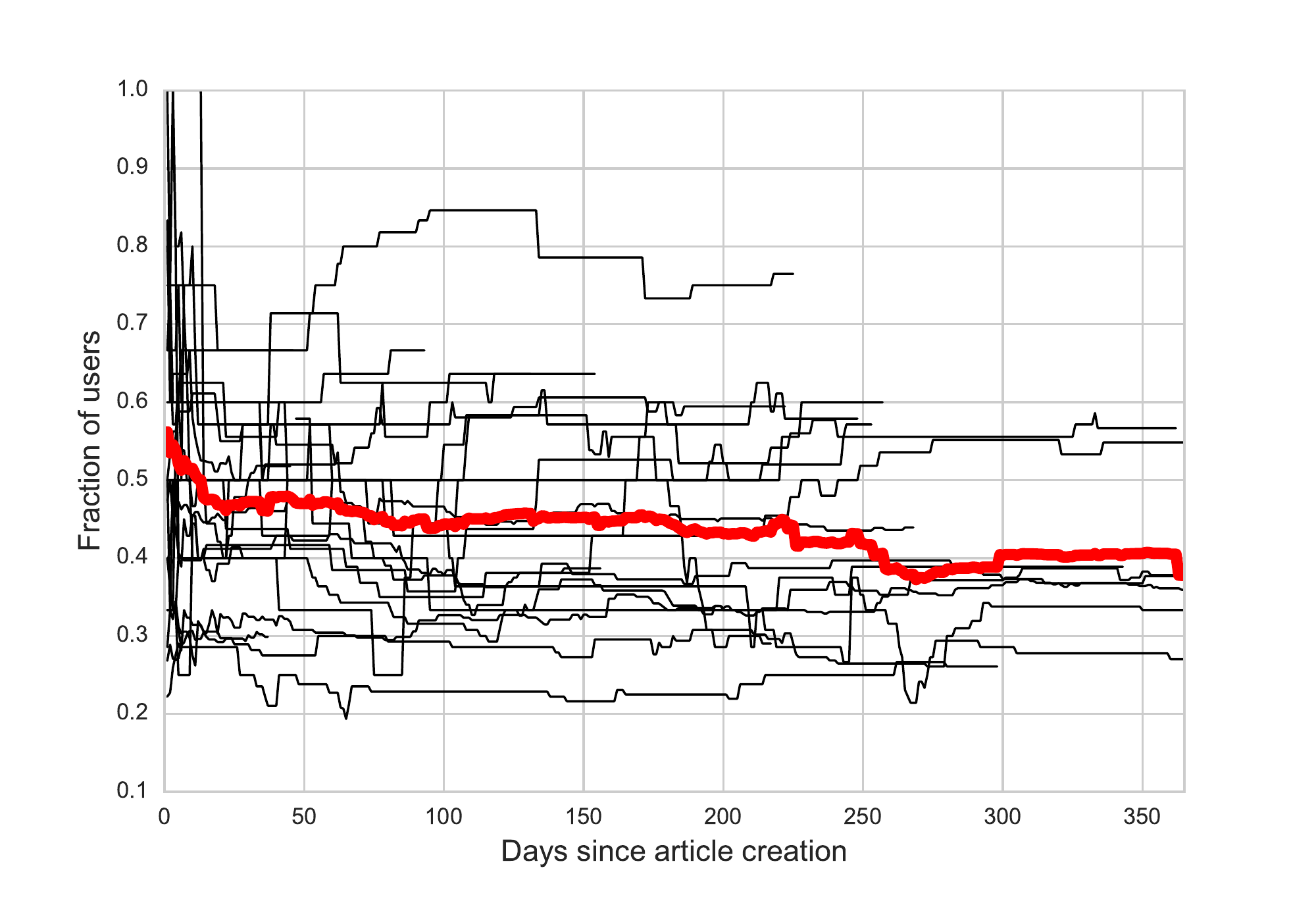}
        \caption{Fraction of editors from Top 10 $BLM$ articles in Table~\ref{tab:rev_activity} collaborating on non-Top 10 $BLM$ articles by day since article creation. Average value across articles in red.}
        \label{fig:top_page_editors}
      \end{figure}

Figure~\ref{fig:top_page_editors} measures editor similarity in an alternative way. It shows the extent to which editors of Top 10 articles (Table~\ref{tab:rev_activity}) collaborated in the rest of the articles over time. We compute the daily fraction of users who edited any of the Top 10 articles and the rest of the articles by \textit{relative dates} since the non-Top 10 article was created. While there is wide variation in articles, the averaged values (in red) show the fraction is initially high ($\sim$50\%) but declines only modestly. Editors migrate throughout the corpus across the lifespans of pages. The tendency for non-Top 10 articles to draw large fractions of their editorship from Top 10 articles, even when many of these articles are created weeks or months after the events themselves, shows that editors collaborate on high-profile events as well as less popular or overlooked events. 
\subsubsection{Black Lives Matter talk page discussion}
The ``Black Lives Matter'' article in Figure~\ref{fig:jaccard_score} has an increasing trend of similarity that differs from other Top 10 articles. The article is the only one in the sample dedicated to the social movement and it is reliant upon the actual events that motivate it. We conducted an interpretive analysis of the ``Black Lives Matter'' talk page to better understand how the events are organized into knowledge on $BLM$. There were 74 editors who made 534 revisions (See Table~\ref{tab:rev_activity}), and 41 of these talk page editors also edited the $BLM$ article itself. This indicates that only 10\% of editors to the article participated in discussions on the talk page. Reviewing the posts that were made from December 21, 2014 to April 3, 2016 reveals deliberations about how Wikipedia rules should be applied to the movement's article.

A recurring debate surrounded $BLM$’s connection to associated events and deaths. Editors deliberated how to situate the $BLM$ article with respect to other related events while maintaining Wikipedia's standards, such as keeping a neutral point of view\footnote{https://en.wikipedia.org/wiki/WP:NPOV}, preventing it from becoming a ``coat rack'' article\footnote{https://en.wikipedia.org/wiki/WP:Coatrack} for enumerating all related events, and identifying reliable sources\footnote{https://en.wikipedia.org/wiki/WP:IRS}. During the time some of these discussions took place, neither the template nor category for $BLM$ existed, and editors discussed whether and how to organize and situate individual events with the movement. One particularly active editor in the sample with 1,040 revisions, ``MrX,'' was adamant that clear connections between related events and $BLM$ needed documentation before inclusion:
\begin{quote}
\textit{Many of the articles linked make no mention of BLM. Again, this is an original research issue and a WP:NPOV issue... My remaining concern is that a list like this may be WP:UNDUE, and turns the article into somewhat of a WP:COATRACK. For example, did Oscar Grant III's death in 2009 actually inspire the movement that started four years later, or is this some revisionist history on the part of the organization?}
\end{quote}

Editors' diligence against such inclusions suggests that Wikipedia remains a place for documenting events and revising accounts of the movement consistent with community standards such as NPOV. This included a debate about identifying a $BLM$ protest. 
The most active editor in the sample, Mandruss (6,248 revisions), expressed concerns about what constitutes a related protest: 
\begin{quote}
\textit{If I then protest the Killing of John Doe and invoke the $BLM$ name, and some [reliable source] reports that I did so (without necessarily endorsing the connection), does that constitute a $BLM$ protest?}
\end{quote}

The talk page illustrates challenges editors faced when trying to produce a neutral and consistent record about a social movement without explicitly supporting it. Identifying reliable sources is challenging with breaking movement events that lack authoritative, centralized sources of information. Editors also needed to decide when to focus content on $BLM$ and to what extent to incorporate pages about individual events into $BLM$ for background. In the context of \textit{collaborative migration}, these discussions show the choices made in organizing knowledge among events before infrastructures (WikiProjects, categories, and templates) exist for topics and are essential for collaboration. 

\subsection{RQ3: Dynamic re-appraisal}
 \begin{figure}[tb]
     \centering
     \includegraphics[width=\columnwidth]{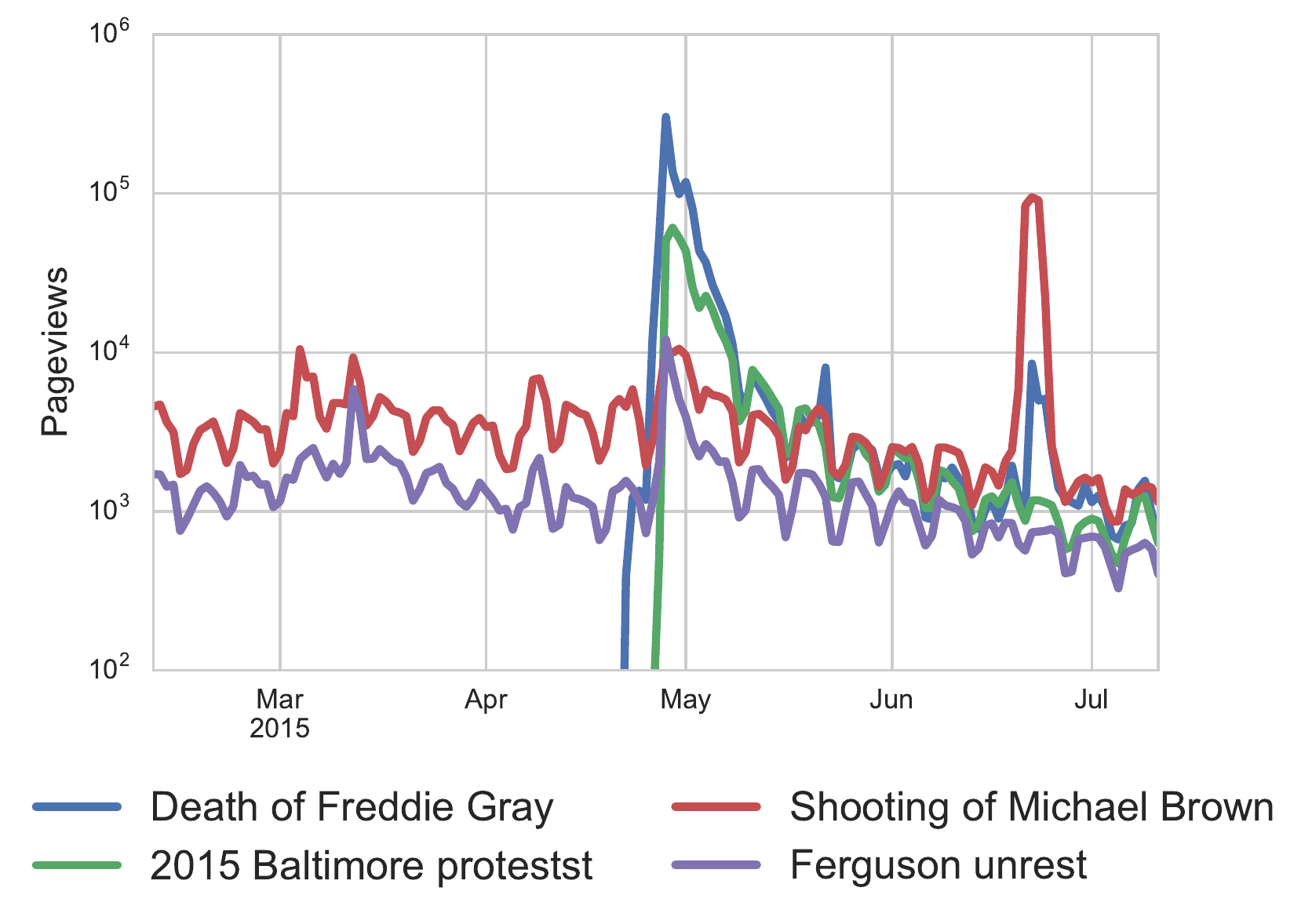}
     \caption{Example: Pageviews among four articles in the aftermath of Freddie Gray's death. Correlations were calculated from the daily attention dynamics.}
     \label{fig:pageview_comparison}
\end{figure}
Research Question 3 asked, ``How are events on Wikipedia re-appraised following new events?'' The declining event-article latency in Figure~\ref{fig:timedelay} suggests relationships between different incidents. Some articles may drive attention or editing activity to other articles. We examine the extent to which revision and pageview behavior about different events are correlated with each other. Unlike other research on social movements, we use these Wikipedia measures to observe how people rely on prior events as part of the sensemaking process associated with a social movement. These relationships highlight a general pattern we term \textit{dynamic re-appraisal} in which Wikipedia users engaged in information seeking return to previous event articles in response to current events.

Figure~\ref{fig:pageview_comparison} is an example of daily pageview activity for four articles in the $BLM$ corpus in the aftermath of Freddie Gray's death in April 2015. The pageviews for all four appear to follow similar trends. There is a spike of activity on the ``Death of Freddie Gray'' and ``2015 Baltimore protests'' articles since these events were major news events at the time. The pageview activity for the ``Shooting of Michael Brown'' and ``Ferguson unrest'' articles also show attention spikes at the same time, despite these events happening nine months earlier. A little over a month later, in late June, we observe a spike in views to the ``Death of Freddie Gray'' and ``Shooting of Michael Brown'' articles without any corresponding increase in views to the ``2015 Baltimore protests'' or ``Ferguson unrest'' articles. These patterns highlight that attention to Wikipedia articles can spillover to adjacent articles~\cite{kummer_spillovers_2014}. We can use such highly-correlated behaviors to reveal topically similar or contextually relevant article relationships.

Tables~\ref{tab:top_rev_corr}--\ref{tab:top_rev_pv_corr} are the most positive correlations for daily revision, pageview, and revision-pageview activity (respectively). Revision activity (Table~\ref{tab:top_rev_corr}) has consistently lower correlations than the pageview (Table~\ref{tab:top_pv_corr}) or revision-pageview relationships (Table~\ref{tab:top_rev_pv_corr}). The cost of editing articles is higher than simply viewing them, which is an intuitive explanation of the observed differences in correlation. The strongest correlations exist between articles about similar events, such as Freddie Gray and Baltimore protests or Michael Brown and Ferguson protests. These correlations also reveal temporal and spatial relationships. The article for Rekia Boyd was created in the same week as the articles for Freddie Gray and 2015 Baltimore protests. The Walter Scott shooting and Charleston church massacre both happened in Charleston, South Carolina in 2015 and the officer's indictment in the former happened days after the latter.

\subsubsection{Clustering of article correlations}
\begin{table}[tb]
 \centering
 \minipage{\columnwidth}
    \centering
        \begin{tabular}{@{}llr@{}}
            \toprule
            Article 1 & Article 2 & Corr. \\
            \midrule
            Freddie Gray & Baltimore protests &  0.672 \\
            Rekia Boyd & Baltimore protests &  0.601 \\
            Jeremy McDole & James Craig Anderson &  0.546 \\
            Tamir Rice & Ferguson unrest &  0.516 \\
            Michael Brown & Baltimore protests &  0.504 \\
            \bottomrule
        \end{tabular}
        \caption{Five article pairs with strongest daily revision correlations.}
        \label{tab:top_rev_corr}
        \vspace{1em}
 \endminipage\hfill\\[1.8ex] 
 \minipage{\columnwidth}
    \centering
        \begin{tabular}{@{}llr@{}}
            \toprule
            Article 1 & Article 2 & Corr. \\
            \midrule
            Michael Brown & Ferguson unrest & 0.918 \\
            Walter Scott & Charleston shooting & 0.915 \\
            Freddie Gray & Baltimore protests & 0.863 \\
            Ferguson unrest & Freddie Gray & 0.842 \\
            Eric Courtney Harris & Freddie Gray & 0.840 \\
            \bottomrule
        \end{tabular}
        \caption{Five article pairs with strongest daily pageview correlations.} 
        \label{tab:top_pv_corr}
        \vspace{1em}
 \endminipage\hfill\\[1.8ex] 
 \minipage{\columnwidth}
    \centering
            \begin{tabular}{@{}llr@{}}
            \toprule
            Pageviews & Revisions & Corr. \\
            \midrule
            Freddie Gray & Baltimore protests &0.942 \\
            Eric Courtney Harris & Baltimore protests &0.856 \\
            Akai Gurley & Baltimore protests &0.839 \\
            Walter Scott & Charleston shooting &0.836 \\
            Ferguson unrest & Baltimore protests &0.829 \\
            \bottomrule
        \end{tabular}
        \caption{Five article pairs with strongest revision-pageview correlations.} 
        \label{tab:top_rev_pv_corr}
  \endminipage
\end{table}

\begin{figure}[t]
    \centering
    \includegraphics[width=\columnwidth]{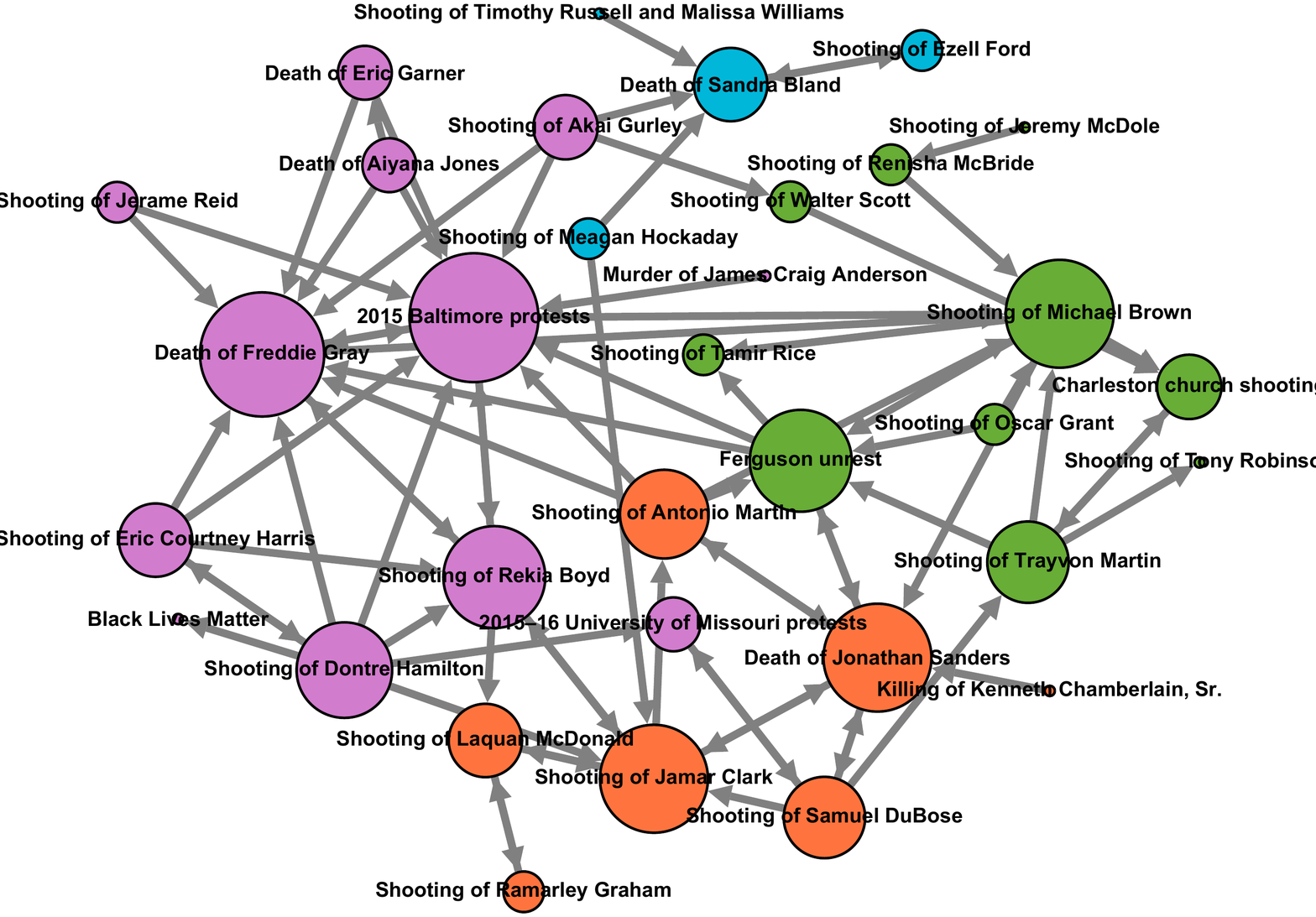}
    \caption{Network of article-article relationships based on correlations among pageview and revision activity. Articles are linked together if these articles occur in the same hierarchical cluster. Nodes are sized by number of connections and colored by a community detection algorithm.}
    \label{fig:cluster_2_hairball}   
\end{figure}
To make sense of these three different correlation relationships, we use hierarchical clustering to identify article pairs that have similar correlation values across all three article interactions. We compute the distances between clusters using the Ward variance minimization algorithm and manually tuned the distance parameter to return four easily-interpretable clusters. The results are clusters where the three correlation values for each article pair in the cluster are more similar to other article pairs in the cluster than outside the cluster. 
Each cluster of article relationships can be qualitatively interpreted as having distinctive information consumption and production patterns. We focus on the cluster where pages' pageview and revisions activity are most strongly and positively correlated with each other. Substantively, these article pairs are viewed and edited in similar ways on similar days---indicating that they have the most well-aligned production and consumption among those in our sample~\cite{warncke-wang_misalignment_2015}.

The set of article relationships in this highly-correlated cluster can also be visualized as a network to identify clusters of topics that behaved most similarly with each other. In Figure~\ref{fig:cluster_2_hairball}, each node in the network is an article and the articles are connected by directed links if these articles' pageview and revision time series behavior appeared in the most strongly correlated cluster. Correlations between articles in this network indicates that both articles' revision and pageview behaviors respond similarly to each other. This network perspective helps to surface latent relationships between topics based on the dynamics of users' activity editing or viewing these pages at the same time. 

This correlation network is not completely connected. Articles about the shootings of Trayvon Martin, Michael Brown, and Oscar Grant, the most prominent and actively-edited events in our corpus, are found in the green-colored community at 3 o'clock in Figure~\ref{fig:cluster_2_hairball}. Despite the tremendous amount of activity on these articles, they are not the most central articles in the correlation network. Articles about the deaths of Freddie Gray and the 2015 Baltimore protests occupy the most central locations in this interaction network (in the pink-colored community at 10 o'clock in Figure~\ref{fig:cluster_2_hairball}).

Activity on these articles drove pageviews and revisions to each other, but not to other articles like ``Black Lives Matter'' (at 9 o'clock in Figure~\ref{fig:cluster_2_hairball}), implying there was little correlation between the ``Black Lives Matter'' article and these prominent event pages. $BLM$ is only a peripheral node in this correlation network, having a single connection to the shooting of Dontre Hamilton. The lack of connectivity for $BLM$ in this cluster implies that the activity on articles about individual events is not robustly correlated with the activity on the $BLM$ movement article. In other words, while articles about different events exhibit strong correlations in pageview and editing activity, this event-related activity is not also driving activity in the article about the movement itself. 

The daily editing and pageviews correlations between related-event articles show that articles can remain active sites of attention even when they are not breaking news stories. These ``dynamic re-appraisal'' processes capture users' engagement with content about some --- but not all --- related events as part of information seeking and sensemaking routines in the aftermath of new events. These behaviors capture emergent knowledge collaboration relationships within social computing systems beyond existing affordances such as coauthorship, hyperlinking, or categorization.

\section{Discussion} 
This analysis of activity around Wikipedia articles related to the \textit{Black Lives Matter} movement contributes to prior literature on social movements by describing knowledge production and collective memory in a social computing system as the movement and related events are happening. We identify three distinct mechanisms --- intensified documentation, collaborative migration, and dynamic re-appraisal --- that occur throughout the $BLM$-related pages on Wikipedia. Through these mechanisms, the articles and the editors who worked on them support collective memory and knowledge production. The activity drives knowledge creation around $BLM$, organizes the topic space, and generates a neutral accounting of the movement and related events. The patterns of participation and attention around $BLM$ pages also demonstrate some novel dynamics when compared with earlier analysis of Wikipedia. We discuss the connections between these mechanisms, theory, the implications of our findings for practice and design, as well as the limitations of this work below. 

\subsection{Intensified documentation}
In the context of the $BLM$ movement, the growth in the number of articles, acceleration of editing activity, and reduction in article creation latency reflect a general pattern of behavior we term \textit{intensified documentation}. 
We do not claim that editors sought to advance the agenda of movement supporters or opponents. Instead, Wikipedians generated a neutral accounting of related events and created topical structures to navigate $BLM$ content. Indirectly, Wikipedia's coverage arguably lends support to $BLM$'s claims about police violence in the United States being a systematic problem rather than isolated cases. This affinity between coverage and movement aims of increased visibility, mourning, and commemoration derive from the specific character and tactics of $BLM$. In general, intensified documentation practices may support or discredit movement frames over time. In this way, intensified Wikipedia coverage of social movement concerns seem likely to capture the context of a movement more thoroughly than other social computing platforms like Twitter.

\subsection{Collaborative migration}
Similar to prior work on breaking news in Wikipedia, we observe \textit{collaborative migrations} as editors work across topically-related articles. In general, effective coverage of new events implies sustained engagement from editors to coordinate in the aftermath, manage conflicts arising in response, and support collaboration by developing infrastructures (like templates and categories)~\cite{keegan_hot_2013}. The \textit{collaborative migrations} around $BLM$ extend the understanding of breaking news in one key way. Wikipedia's coverage of $BLM$ requires the integration of the social movement with events across a longer time horizon. Some events are breaking news with respect to the movement, but others are older events. Some events had articles well \textit{before} $BLM$ existed, while other events occurred years ago, but did not have an article until \textit{after} $BLM$. New current event articles also prompted noticeable changes in the similarities of existing articles, sometimes increasing similarity as prior editors came in or decreasing similarity as new editors joined the editing population. The $BLM$ article itself did not have high co-authorship similarity, but exhibited sustained growth in similarity unlike the event articles. This suggests that it played an increasingly central role facilitating collaboration and connections across events as its editor population became more similar to the other articles. There are likely other types of current events and topic domains that fit this pattern of a central reference article surrounded by a set of less connected articles.

\subsection{Dynamic re-appraisal}

The re-appraisal of past events in response to new events highlights how the distinct affordances of Wikipedia translate into collective memory practices. 
We find evidence of increased traffic to articles about past events even when there was little new information on them. We interpret this as evidence of Wikipedia readers making sense of new events by reading about related prior events. We are not aware of prior work documenting this pattern on Wikipedia or other social computing systems.

We also observe correlations between knowledge production (revisions) and consumption (pageviews) within the same article and also between different articles in our study. In the context of $BLM$ these correlations potentially reflect collective sense-making attempts to constrain uncertain, anomalous, or unjust events within narratives about events having similar familiar contexts and actors~\cite{polletta_contending_1998}. They also reflect patterns consistent with past Wikipedia research finding correlations between viewership and editorship \cite{hill_consider_2014}.

Intentionally or not, the process of dynamic re-appraisal around events of public mourning and commemoration aligns with some of the goals and tactics embraced by $BLM$ movement participants. 
Without breaking with community norms like NPOV, Wikipedia became a site of collective memory documenting mourning practices as well as tracing how memories were encoded and re-interpreted. Editors and readers who curated, visited, and re-visited information surrounding movement-related events contributed to this process.

\subsection{Implications for practice and design}
Movement participants or organizations may look to the $BLM$ coverage on Wikipedia as an example of how to coopt a social computing system, but this reflects a profound misunderstanding of how Wikipedia works. Wikipedia's open nature and encyclopedic scope allow social movements to be neutrally documented, resulting in framing and narrative generation processes unlike those found in other social computing systems. Typically, a movement's framing describes the grievances at stake, determines how the movement is understood, and specifies the scope of issues to be addressed by fostering a ``sense of severity, urgency, efficacy, and propriety''~\cite{benford_you_1993}. Movement narratives attempt to link past, present, and future events together and align individual and collective identities into stories that shared and referenced~\cite{polletta_contending_1998}. While the use of social media has transformed mobilization strategies~\cite{bennett_logic_2013, earl_digitally_2011}, movement actors should also consider the role of online communities like Wikipedia in framing movements for a broader audience. Distinct from settings like Twitter or Facebook, movements may struggle to exert control over Wikipedia narratives since the information must meet criteria like neutrality and notability.

Designers could better support the sorts of collaboration, collective sense-making, and attention dynamics we observed around $BLM$ articles in Wikipedia. For example, the evidence we uncover of dynamic re-appraisal suggests that editors working on Wikipedia articles about a particular current event would likely benefit from understanding the coverage of related prior events. Likewise, readership patterns, category structures, and templates could be used to design a more active content suggestion system for Wikipedia readers. Previous work indicates that surfacing past activities or interests for editors can mobilize them to make more contributions~\cite{cosley_suggestbot_2007}. Our findings here imply that similar interventions could help readers navigate a topically-linked set of articles related to unfolding events or movements.

\subsection{Limitations and future work}
Our analysis looked at only a single, prominent, contemporary social movement case in the United States and activity related to it on the largest language edition of Wikipedia. All of these features potentially limit the generalizability of these findings across other cases, movements, or contexts. However, social movements are ubiquitous and social computing systems increasingly function as spaces of collective knowledge production, sense-making, and commemoration. The dynamics we have described here can help guide future investigations into these phenomena.


A richer mixed-methods accounting of the framing contests, value conflicts, editor identities, and motivations would also complement the results we report. The logged revisions provide a rich, but convenient sample of data that censors contributions during page protections~\cite{hill_page_2015}, overlooks the choice of information sources~\cite{ford_wikipedia_2012,ford_getting_2013} and the role of automated agents in supporting collaboration~\cite{geiger_when_2013}. We do not have data on the demographic attributes of editors or readers in our dataset. While prior work has documented disparities in participation~\cite{hargittai_mind_2015}, we do not know whether or how the readers and editors of pages related to $BLM$ reflect these broader trends. In addition, while the representation of $BLM$ in Wikipedia may advance some movement goals, we cannot make claims regarding whether editors support the movement or not. Future research might address this in a variety of ways, including analyzing the content of revisions and revision histories of editors, administering surveys to collect demographic or motivational data, and performing interviews and qualitative analyses of editors' actions. 


The descriptive and exploratory quantitative analyses we employed could be augmented as well. Regression, network, or time series modeling could be used to estimate the likelihood of existing editors contributing to movement articles or editors migrating across articles. Text analysis techniques could be used in conjunction with news data to examine the gaps or biases in coverage on Wikipedia compared to popular media narratives and frames. Sequence analysis approaches could be used to better disambiguate the direction of editor flows among these articles~\cite{keegan_analyzing_2016}. Observational causal inference approaches could also be used to identify the influence of dynamic re-appraisals on other collaborative proceses~\cite{kummer_spillovers_2014}. 

\section{Conclusion}
Wikipedia's openness, popularity, and legibility provides a unique opportunity to track the evolution of social movements' activity over time. Our analysis of the English Wikipedia's response to the ``Black Lives Matter'' movement and articles about related events extends understanding of the role social computing systems have in online collective action. Our findings point to the interplay between collective action and collective memory by highlighting three distinctive mechanisms: intensified documentation, collaborative migration, and dynamic re-appraisal.

\section{Acknowledgements}
The authors thank the BYOR workshop at Northwestern for reviewing early drafts of this work. We thank the anonymous reviewers and Cliff Lampe for their detailed comments, which have strengthened this work.
\balance
\bibliographystyle{SIGCHI-Reference-Format} 
\bibliography{refs}


\begin{thebibliography}{00}


\ifx \showCODEN    \undefined \def \showCODEN     #1{\unskip}     \fi
\ifx \showDOI      \undefined \def \showDOI       #1{{\tt DOI:}\penalty0{#1}\ }
  \fi
\ifx \showISBNx    \undefined \def \showISBNx     #1{\unskip}     \fi
\ifx \showISBNxiii \undefined \def \showISBNxiii  #1{\unskip}     \fi
\ifx \showISSN     \undefined \def \showISSN      #1{\unskip}     \fi
\ifx \showLCCN     \undefined \def \showLCCN      #1{\unskip}     \fi
\ifx \shownote     \undefined \def \shownote      #1{#1}          \fi
\ifx \showarticletitle \undefined \def \showarticletitle #1{#1}   \fi
\ifx \showURL      \undefined \def \showURL       #1{#1}          \fi

\bibitem{arazy_functional_2015}
{Ofer Arazy}, {Felipe Ortega}, {Oded Nov}, {Lisa Yeo}, {and} {Adam Balila}.
  2015.
\newblock \showarticletitle{Functional {Roles} and {Career} {Paths} in
  {Wikipedia}}. In {\em Proceedings of the 18th {ACM} {Conference} on
  {Computer} {Supported} {Cooperative} {Work} \& {Social} {Computing}} {\em
  ({CSCW} '15)}. ACM, New York, NY, USA, 1092--1105.
\newblock
\showISBNx{978-1-4503-2922-4}
\showDOI{%
\url{http://dx.doi.org/10.1145/2675133.2675257}}


\bibitem{benford_you_1993}
{Robert~D Benford}. 1993.
\newblock \showarticletitle{“{You} {Could} {Be} the {Hundredth} {Monkey}”:
  {Collective} {Action} {Frames} and {Vocabularies} of {Motive} {Within} the
  {Nuclear} {Disarmament} {Movement}}.
\newblock {\em The Sociological Quarterly\/} {34}, 2 (1993), 195--216.
\newblock


\bibitem{bennett_logic_2013}
{W.~Lance Bennett} {and} {Alexandra Segerberg}. 2013.
\newblock {\em The {Logic} of {Connective} {Action}: {Digital} {Media} and the
  {Personalization} of {Contentious} {Politics}}.
\newblock Cambridge University Press, New York.
\newblock
\showISBNx{978-1-107-02574-5}


\bibitem{cosley_suggestbot_2007}
{Dan Cosley}, {Dan Frankowski}, {Loren Terveen}, {and} {John Riedl}. 2007.
\newblock \showarticletitle{SuggestBot: Using Intelligent Task Routing to Help
  People Find Work in Wikipedia}. In {\em Proceedings of the 12th International
  Conference on Intelligent User Interfaces} {\em (IUI '07)}. ACM, New York,
  NY, USA, 32--41.
\newblock
\showISBNx{1-59593-481-2}
\showDOI{%
\url{http://dx.doi.org/10.1145/1216295.1216309}}


\bibitem{de_choudhury_social_2016}
{Munmun De~Choudhury}, {Shagun Jhaver}, {Benjamin Sugar}, {and} {Ingmar Weber}.
  2016.
\newblock \showarticletitle{Social {Media} {Participation} in an {Activist}
  {Movement} for {Racial} {Equality}}. In {\em Tenth {International} {AAAI}
  {Conference} on {Web} and {Social} {Media}}.
\newblock
\showURL{%
\url{http://www.aaai.org/ocs/index.php/ICWSM/ICWSM16/paper/view/13168}}


\bibitem{earl_digitally_2011}
{Jennifer Earl} {and} {Katrina Kimport}. 2011.
\newblock {\em Digitally {Enabled} {Social} {Change}: {Activism} in the
  {Internet} {Age}}.
\newblock MIT Press.
\newblock
\showISBNx{978-0-262-01510-3}


\bibitem{ferron_collective_2011}
{Michela Ferron} {and} {Paolo Massa}. 2011.
\newblock \showarticletitle{Collective memory building in {Wikipedia}: the case
  of {North} {African} uprisings}.
\newblock {\em Proceedings of the 7th International Symposium on Wikis and Open
  Collaboration\/} (2011), 114--123.
\newblock
\showISSN{1450309097}


\bibitem{ford_wikipedia_2012}
{Heather Ford}. 2012.
\newblock {\em Wikipedia {Sources}: {Managing} {Sources} in {Rapidly}
  {Evolving} {Global} {News} {Articles} on the {English} {Wikipedia}}.
\newblock {SSRN} {Scholarly} {Paper} ID 2127204. Social Science Research
  Network, Rochester, NY.
\newblock
\showURL{%
\url{http://papers.ssrn.com/abstract=2127204}}


\bibitem{ford_getting_2013}
{Heather Ford}, {Shilad Sen}, {David~R. Musicant}, {and} {Nathaniel Miller}.
  2013.
\newblock \showarticletitle{Getting to the {Source}: {Where} {Does} {Wikipedia}
  {Get} {Its} {Information} from?}. In {\em Proceedings of the 9th
  {International} {Symposium} on {Open} {Collaboration}} {\em ({WikiSym} '13)}.
  ACM, New York, NY, USA, 9:1--9:10.
\newblock
\showISBNx{978-1-4503-1852-5}
\showDOI{%
\url{http://dx.doi.org/10.1145/2491055.2491064}}


\bibitem{freelon_beyond_2016}
{Deen~Goodwin Freelon}, {Charlton~D. McIlwain}, {and} {Meredith~D. Clark}.
  2016.
\newblock \showarticletitle{Beyond the {Hashtags}:\# {Ferguson},\#
  {Blacklivesmatter}, and the {Online} {Struggle} for {Offline} {Justice}}.
\newblock {\em SSRN eLibrary\/} (2016).
\newblock
\showURL{%
\url{http://papers.ssrn.com/sol3/papers.cfm?abstract_id=2747066}}


\bibitem{geiger_when_2013}
{R.~Stuart Geiger} {and} {Aaron Halfaker}. 2013.
\newblock \showarticletitle{When the {Levee} {Breaks}: {Without} {Bots}, {What}
  {Happens} to {Wikipedia}'s {Quality} {Control} {Processes}?}. In {\em
  Proceedings of the 9th {International} {Symposium} on {Open} {Collaboration}}
  {\em ({WikiSym} '13)}. ACM, New York, NY, USA, 6:1--6:6.
\newblock
\showISBNx{978-1-4503-1852-5}
\showDOI{%
\url{http://dx.doi.org/10.1145/2491055.2491061}}
\newblock
\shownote{00010.}


\bibitem{gongaware_collective_2010}
{Timothy~B Gongaware}. 2010.
\newblock \showarticletitle{Collective memory anchors: collective identity and
  continuity in social movements}.
\newblock {\em Sociological Focus\/} {43}, 3 (2010), 214--239.
\newblock


\bibitem{graeff_battle_2014}
{Erhardt Graeff}, {Matt Stempeck}, {and} {Ethan Zuckerman}. 2014.
\newblock \showarticletitle{The battle for ‘{Trayvon} {Martin}’: {Mapping}
  a media controversy online and off-line}.
\newblock {\em First Monday\/} {19}, 2 (Jan. 2014).
\newblock
\showISSN{13960466}
\showURL{%
\url{http://firstmonday.org/ojs/index.php/fm/article/view/4947}}


\bibitem{halbwachs_collective_1992}
{Maurice Halbwachs} {and} {Lewis~A Coser}. 1992.
\newblock {\em On collective memory}.
\newblock University of Chicago Press.
\newblock
\showISBNx{0-226-11596-8}


\bibitem{hargittai_mind_2015}
{Eszter Hargittai} {and} {Aaron Shaw}. 2015.
\newblock \showarticletitle{Mind the skills gap: the role of {Internet}
  know-how and gender in differentiated contributions to {Wikipedia}}.
\newblock {\em Information, Communication \& Society\/} {18}, 4 (April 2015),
  424--442.
\newblock
\showISSN{1369-118X}
\showDOI{%
\url{http://dx.doi.org/10.1080/1369118X.2014.957711}}


\bibitem{harris_it_2006}
{Fredrick~C Harris}. 2006.
\newblock \showarticletitle{It takes a tragedy to arouse them: {Collective}
  memory and collective action during the civil rights movement}.
\newblock {\em Social movement studies\/} {5}, 1 (2006), 19--43.
\newblock


\bibitem{hill_consider_2014}
{Benjamin~Mako Hill} {and} {Aaron Shaw}. 2014.
\newblock \showarticletitle{Consider the {Redirect}: {A} {Missing} {Dimension}
  of {Wikipedia} {Research}}. In {\em Proceedings of {The} {International}
  {Symposium} on {Open} {Collaboration}} {\em ({OpenSym} '14)}. ACM, New York,
  NY, USA, 28:1--28:4.
\newblock
\showISBNx{978-1-4503-3016-9}
\showDOI{%
\url{http://dx.doi.org/10.1145/2641580.2641616}}
\newblock
\shownote{00002.}


\bibitem{hill_page_2015}
{Benjamin~Mako Hill} {and} {Aaron Shaw}. 2015.
\newblock \showarticletitle{Page {Protection}: {Another} {Missing} {Dimension}
  of {Wikipedia} {Research}}. In {\em Proceedings of the 11th {International}
  {Symposium} on {Open} {Collaboration}} {\em ({OpenSym} '15)}. ACM, New York,
  NY, USA, 15:1--15:4.
\newblock
\showISBNx{978-1-4503-3666-6}
\showDOI{%
\url{http://dx.doi.org/10.1145/2788993.2789846}}


\bibitem{jackson__2015}
{Sarah~J Jackson} {and} {Brooke Foucault~Welles}. 2015a.
\newblock \showarticletitle{\# {Ferguson} is everywhere: initiators in emerging
  counterpublic networks}.
\newblock {\em Information, Communication \& Society\/} (2015), 1--22.
\newblock
\showISSN{1369-118X}


\bibitem{jackson_hijacking_2015}
{Sarah~J. Jackson} {and} {Brooke Foucault~Welles}. 2015b.
\newblock \showarticletitle{Hijacking \#{myNYPD}: {Social} {Media} {Dissent}
  and {Networked} {Counterpublics}}.
\newblock {\em Journal of Communication\/} {65}, 6 (Dec. 2015), 932--952.
\newblock
\showISSN{1460-2466}
\showDOI{%
\url{http://dx.doi.org/10.1111/jcom.12185}}


\bibitem{jemielniak_common_2014}
{Dariusz Jemielniak}. 2014.
\newblock {\em Common {Knowledge}?: {An} {Ethnography} of {Wikipedia}}.
\newblock Stanford University Press, Stanford, California.
\newblock
\showISBNx{978-0-8047-8944-8}


\bibitem{keegan_history_2013}
{Brian~C. Keegan}. 2013.
\newblock \showarticletitle{A {History} of {Newswork} on {Wikipedia}}. In {\em
  Proceedings of the 9th {International} {Symposium} on {Open} {Collaboration}}
  {\em ({WikiSym} '13)}. ACM, New York, NY, USA, 7:1--7:10.
\newblock
\showISBNx{978-1-4503-1852-5}
\showDOI{%
\url{http://dx.doi.org/10.1145/2491055.2491062}}


\bibitem{keegan_emergent_2015}
{Brian~C Keegan}. 2015.
\newblock \showarticletitle{Emergent {Social} {Roles} in {Wikipedia}’s
  {Breaking} {News} {Collaborations}}.
\newblock In {\em Roles, {Trust}, and {Reputation} in {Social} {Media}
  {Knowledge} {Markets}}. Springer, 57--79.
\newblock


\bibitem{keegan_is_2015}
{Brian~C. Keegan} {and} {Jed~R. Brubaker}. 2015.
\newblock \showarticletitle{'{Is}' to '{Was}': {Coordination} and
  {Commemoration} in {Posthumous} {Activity} on {Wikipedia} {Biographies}}. In
  {\em Proceedings of the 18th {ACM} {Conference} on {Computer} {Supported}
  {Cooperative} {Work} \& {Social} {Computing}} {\em ({CSCW} '15)}. ACM, New
  York, NY, USA, 533--546.
\newblock
\showISBNx{978-1-4503-2922-4}
\showDOI{%
\url{http://dx.doi.org/10.1145/2675133.2675238}}


\bibitem{keegan_hot_2011}
{Brian~C. Keegan}, {Darren Gergle}, {and} {Noshir Contractor}. 2011.
\newblock \showarticletitle{Hot off the wiki: dynamics, practices, and
  structures in {Wikipedia}'s coverage of the {T\={o}hoku} catastrophes}.
\newblock {\em Proceedings of the 7th international symposium on Wikis and open
  collaboration\/} (2011), 105--113.
\newblock
\showISSN{1450309097}


\bibitem{keegan_staying_2012}
{Brian~C. Keegan}, {Darren Gergle}, {and} {Noshir Contractor}. 2012.
\newblock \showarticletitle{Staying in the {Loop}: {Structure} and {Dynamics}
  of {Wikipedia}'s {Breaking} {News} {Collaborations}}. In {\em Proceedings of
  the {Eighth} {Annual} {International} {Symposium} on {Wikis} and {Open}
  {Collaboration}} {\em ({WikiSym} '12)}. ACM, New York, NY, USA, 1:1--1:10.
\newblock
\showISBNx{978-1-4503-1605-7}
\showDOI{%
\url{http://dx.doi.org/10.1145/2462932.2462934}}


\bibitem{keegan_hot_2013}
{Brian~C. Keegan}, {Darren Gergle}, {and} {Noshir Contractor}. 2013.
\newblock \showarticletitle{Hot {Off} the {Wiki} {Structures} and {Dynamics} of
  {Wikipedia}'s {Coverage} of {Breaking} {News} {Events}}.
\newblock {\em American Behavioral Scientist\/} {57}, 5 (May 2013), 595--622.
\newblock
\showISSN{0002-7642, 1552-3381}
\showDOI{%
\url{http://dx.doi.org/10.1177/0002764212469367}}


\bibitem{keegan_analyzing_2016}
{Brian~C. Keegan}, {Shakked Lev}, {and} {Ofer Arazy}. 2016.
\newblock \showarticletitle{Analyzing {Organizational} {Routines} in {Online}
  {Knowledge} {Collaborations}: {A} {Case} for {Sequence} {Analysis} in
  {CSCW}}. In {\em Proceedings of the 19th {ACM} {Conference} on
  {Computer}-{Supported} {Cooperative} {Work} \& {Social} {Computing}} {\em
  ({CSCW} '16)}. ACM, New York, NY, USA, 1065--1079.
\newblock
\showISBNx{978-1-4503-3592-8}
\showDOI{%
\url{http://dx.doi.org/10.1145/2818048.2819962}}


\bibitem{kittur_harnessing_2008}
{Aniket Kittur} {and} {Robert~E. Kraut}. 2008.
\newblock \showarticletitle{Harnessing the {Wisdom} of {Crowds} in {Wikipedia}:
  {Quality} {Through} {Coordination}}. In {\em Proceedings of the 2008 {ACM}
  {Conference} on {Computer} {Supported} {Cooperative} {Work}} {\em ({CSCW}
  '08)}. ACM, New York, NY, USA, 37--46.
\newblock
\showISBNx{978-1-60558-007-4}
\showDOI{%
\url{http://dx.doi.org/10.1145/1460563.1460572}}


\bibitem{kittur_he_2007}
{Aniket Kittur}, {Bongwon Suh}, {Bryan~A. Pendleton}, {and} {Ed~H. Chi}. 2007.
\newblock \showarticletitle{He {Says}, {She} {Says}: {Conflict} and
  {Coordination} in {Wikipedia}}. In {\em Proceedings of the {SIGCHI}
  {Conference} on {Human} {Factors} in {Computing} {Systems}} {\em ({CHI}
  '07)}. ACM, New York, NY, USA, 453--462.
\newblock
\showISBNx{978-1-59593-593-9}
\showDOI{%
\url{http://dx.doi.org/10.1145/1240624.1240698}}


\bibitem{konieczny_governance_2009}
{Piotr Konieczny}. 2009.
\newblock \showarticletitle{Governance, {Organization}, and {Democracy} on the
  {Internet}: {The} {Iron} {Law} and the {Evolution} of {Wikipedia}}.
\newblock {\em Sociological Forum\/} {24}, 1 (2009), 162--192.
\newblock
\showISSN{1573-7861}
\showDOI{%
\url{http://dx.doi.org/10.1111/j.1573-7861.2008.01090.x}}


\bibitem{kummer_spillovers_2014}
{Michael~E. Kummer}. 2014.
\newblock {\em Spillovers in {Networks} of {User} {Generated} {Content}:
  {Pseudo}-{Experimental} {Evidence} on {Wikipedia}}.
\newblock {SSRN} {Scholarly} {Paper} ID 2567179. Social Science Research
  Network, Rochester, NY.
\newblock
\showURL{%
\url{http://papers.ssrn.com/abstract=2567179}}


\bibitem{luyt_wikipedia_2015}
{Brendan Luyt}. 2015.
\newblock \showarticletitle{Wikipedia, collective memory, and the {Vietnam}
  war}.
\newblock {\em Journal of the Association for Information Science and
  Technology\/} (2015).
\newblock
\showISSN{2330-1643}


\bibitem{mcadam_recruitment_1986}
{Doug McAdam}. 1986.
\newblock \showarticletitle{Recruitment to high-risk activism: {The} case of
  freedom summer}.
\newblock {\em American journal of sociology\/} (1986), 64--90.
\newblock
\showISSN{0002-9602}


\bibitem{meraz_networked_2013}
{Sharon Meraz} {and} {Zizi Papacharissi}. 2013.
\newblock \showarticletitle{Networked {Gatekeeping} and {Networked} {Framing}
  on \#{Egypt}}.
\newblock {\em The International Journal of Press/Politics\/} (Jan. 2013),
  138--166.
\newblock
\showISSN{1940-1612, 1940-1620}
\showDOI{%
\url{http://dx.doi.org/10.1177/1940161212474472}}


\bibitem{pentzold_fixing_2009}
{Christian Pentzold}. 2009.
\newblock \showarticletitle{Fixing the floating gap: {The} online encyclopaedia
  {Wikipedia} as a global memory place}.
\newblock {\em Memory Studies\/} {2}, 2 (2009), 255--272.
\newblock
\showISSN{1750-6980}


\bibitem{polletta_contending_1998}
{Francesca Polletta}. 1998.
\newblock \showarticletitle{Contending stories: {Narrative} in social
  movements}.
\newblock {\em Qualitative Sociology\/} {21}, 4 (1998), 419--446.
\newblock


\bibitem{salamon_fear_1973}
{Lester~M Salamon} {and} {Stephen Van~Evera}. 1973.
\newblock \showarticletitle{Fear, apathy, and discrimination: {A} test of three
  explanations of political participation}.
\newblock {\em American Political Science Review\/} {67}, 04 (1973),
  1288--1306.
\newblock


\bibitem{schuman_generations_1989}
{Howard Schuman} {and} {Jacqueline Scott}. 1989.
\newblock \showarticletitle{Generations and collective memories}.
\newblock {\em American sociological review\/} (1989), 359--381.
\newblock


\bibitem{sharma_black_2013}
{Sanjay Sharma}. 2013.
\newblock \showarticletitle{Black {Twitter}?: {Racial} hashtags, networks and
  contagion}.
\newblock {\em New formations: a journal of culture/theory/politics\/} {78}, 1
  (2013), 46--64.
\newblock
\showISSN{1741-0789}


\bibitem{tufekci_social_2012}
{Zeynep Tufekci} {and} {Christopher Wilson}. 2012.
\newblock \showarticletitle{Social {Media} and the {Decision} to {Participate}
  in {Political} {Protest}: {Observations} {From} {Tahrir} {Square}}.
\newblock {\em Journal of Communication\/} {62}, 2 (2012), 363--379.
\newblock
\showISSN{1460-2466}
\showDOI{%
\url{http://dx.doi.org/10.1111/j.1460-2466.2012.01629.x}}


\bibitem{warncke-wang_misalignment_2015}
{Morten Warncke-Wang}, {Vivek Ranjan}, {Loren Terveen}, {and} {Brent Hecht}.
  2015.
\newblock \showarticletitle{Misalignment {Between} {Supply} and {Demand} of
  {Quality} {Content} in {Peer} {Production} {Communities}}.
\newblock {\em ICWSM 2015: Ninth International AAAI Conference on Web and
  Social Media\/} (2015).
\newblock
\showURL{%
\url{http://www-users.cs.umn.edu/~morten/publications/icwsm2015-popularity-quality-misalignment.pdf}}


\bibitem{welser_finding_2011}
{Howard~T. Welser}, {Dan Cosley}, {Gueorgi Kossinets}, {Austin Lin}, {Fedor
  Dokshin}, {Geri Gay}, {and} {Marc Smith}. 2011.
\newblock \showarticletitle{Finding social roles in {Wikipedia}}. In {\em
  Proceedings of the 2011 {iConference}} {\em ({iConference} '11)}. ACM, New
  York, NY, USA, 122--129.
\newblock
\showISBNx{978-1-4503-0121-3}
\showDOI{%
\url{http://dx.doi.org/10.1145/1940761.1940778}}


\bibitem{yasseri_dynamics_2012}
{Taha Yasseri}, {Robert Sumi}, {Andras Rung}, {Andras Kornai}, {and} {János
  Kertesz}. 2012.
\newblock \showarticletitle{Dynamics of {Conflicts} in {Wikipedia}}.
\newblock {\em PLoS ONE\/} {7}, 6 (June 2012), e38869.
\newblock
\showDOI{%
\url{http://dx.doi.org/10.1371/journal.pone.0038869}}


\bibitem{zhang_wedo:_2014}
{Haoqi Zhang}, {Andrés Monroy-Hernández}, {Aaron Shaw}, {Sean~A. Munson},
  {Elizabeth Gerber}, {Benjamin~Mako Hill}, {Peter Kinnaird}, {Shelly~D.
  Farnham}, {and} {Patrick Minder}. 2014.
\newblock \showarticletitle{{WeDo}: {End}-{To}-{End} {Computer} {Supported}
  {Collective} {Action}}. In {\em Eighth {International} {AAAI} {Conference} on
  {Weblogs} and {Social} {Media}}.
\newblock
\showURL{%
\url{http://www.aaai.org/ocs/index.php/ICWSM/ICWSM14/paper/view/8041}}


\end{thebibliography}
\end{document}